\input epsf                                                       %
\documentstyle[12pt]{article}
\begin{document}
\title{INFORMATION IN BLACK HOLE  RADIATION\thanks
{Alberta-Thy-24-93, hep-th/9306083.}}
\author{ Don N. Page\\
CIAR Cosmology Program\\
Theoretical Physics Institute\\
Department of Physics\\University of Alberta\\
Edmonton, Alberta\\Canada T6G 2J1\\
don@phys.ualberta.ca}
\date{(1993 June 17; revised Aug. 24)}
\maketitle
\large
\begin{abstract}
\baselineskip 16pt
     If black hole formation and evaporation can be described by an
$S$ matrix, information would be expected to come out in black hole
radiation.  An estimate shows that it
may come out initially so slowly, or else be so spread out,
that it would never show up in an analysis
perturbative in $M_{Planck}/M$, or in $1/N$ for
two-dimensional dilatonic black holes with a large number
$N$ of minimally coupled scalar fields.
\\
\\
\end{abstract}
\normalsize
\pagebreak
\baselineskip 15pt

	Hawking's calculation of thermal emission from a
stationary classical black hole \cite{Haw74,Haw76}
soon led to a major unresolved puzzle concerning quantum mechanics
and gravity:  what happens to a pure quantum state that collapses
to form a black hole which emits approximately themal radiation?
Hawking proposed \cite{Haw76} that the black hole would eventually
disappear completely and that the resulting state of radiation,
like a precisely thermal state, would be mixed.  In other words,
information would be permanently lost down the black hole,
and there would be no $S$ matrix to take an initial pure state
to a final pure state.

	    It was soon objected \cite{Zel77b,Pag80} that this
conclusion
is not justified by the classical or semiclassical approximation for
the
black hole used to derive it, and that, in its original form at
least, it
violates a strong form of $CPT$ invariance \cite{Pag80}.  A number of
alternative possibilities were given \cite{Pag80}.  The main options
now under active investigation seem to be that either most of the
information comes out with the bulk of the radiation to give an $S$
matrix
\cite{Pag80,Hoo,uni2DBH,STU}, or most of the information goes
into a long-lived \cite{ACN,Pre92} or absolutely stable remnant
\cite{BDDO'L}, or else information
is lost from our universe as Hawking proposed \cite{Haw76}.
For recent reviews of the problem, see
\cite{Pre92,HS,Gid92,Pag93,DanSch,Gid93}.

	The first of these options is in some sense the most
conservative,
and I have advocated it as the most productive to pursue
\cite{Pag80}.
However, a number of arguments have been given against this
possibility.  Hawking's original proposal of a loss of information
\cite{Haw76} argued that the semiclassical approximation (which
would give a loss of information) is valid until the black hole gets
down near the Planck mass, and then there is not enough energy left
to carry the information (implicitly assuming no long-lived or stable
remnants).  Giddings and Nelson \cite{GidNel}, and later Giddings
alone \cite{Gid92}, gave a more detailed version of this argument
for a more tractable model of two-dimensional dilatonic black holes
with $N$ minimally coupled scalar fields \cite{CGHS}.

	In this two-dimensional model, the classical equations can be
solved exactly \cite{CGHS}, and the semiclassical equations can be
solved numerically \cite {Pir93}, though a full quantum solution for
this model or any realistic variant is still out of reach.  The
semiclassical
approximation appears to be good until the black hole reaches the
strong-coupling regime (the analogue of the Planck mass for
four-dimensional black holes), by which time the black hole has
emitted most of its energy if its initial thermodynamic entropy
$s_h$ is large compared to $N$.  At least the semiclassical
analysis seems to be valid until then for certain aspects
of the problem, such as the average emission rate,
though perhaps not for the information, as I shall argue below.

	Assuming the validity of the semiclassical analysis until the
weak-coupling approximation breaks down, Giddings and Nelson
\cite{GidNel} conclude, ``The above arguments therefore strongly
suggest that within the present model information does not escape
until the black hole is very small.  Making these rigorous will
therefore
rule out one suggested resolution of the black-hole information
problem, namely, that the information escapes over the course of
black-hole evaporation if the effects of the back reaction are
included"
\cite{GidNel}.  Giddings later stated this more cautiously, that
``working order-by-order in $1/N$, it is probable that
one can construct an argument...analogous to stating that the
information
doesn't come out of four-dimensional black holes until they reach the
Planck
scale" \cite{Gid92}.

	Here I wish to object that if the information does indeed
come out
gradually over the entire emission process, it appears likely that
the rate
of information outflow may initially be so low that it would not show
up
in an order-by-order (perturbative) analysis, and that the
information
in the entire emission would be so spread out that it would require
too
many measurements to be found or excluded by a perturbative analysis.

	An extreme example that shows how this is as least
theoretically
possible is the two-dimensional moving mirror model analyzed by
Carlitz and Willey and by Wilczek \cite{CW2}.  In this model
the early Hawking radiation is {\it exactly} thermal, in a maximally
mixed
state with no information, but then is entirely correlated with the
late
radiation, so that the total state of all the radiation is pure,
containing
all the initial information.  An order-by-order analysis of the early
radiation would reveal no information, but the tempting conclusion
that
the information cannot escape until the black hole gets near the
Planck
mass would be invalidated by the nonanalytic change in the
information
rate at the beginning of the late radiation that is correlated with
the
early radiation.  Only by accurately measuring a huge number of
correlations between the early and late radiation could one hope
to find the information.

	One might counter that this extreme case of exactly thermal
local
radiation, with correlations only between the first half and the
second
half, is not at all plausible.  Therefore, here I shall here examine
a more
natural model, in which the black hole and its surrounding radiation
are two subsystems of a combined system which is assumed to be in
a random pure state.  Tracing over the black hole subsystem gives a
statistical state (i.e., a state represented by a density matrix) for
the
radiation subsystem that generically is mixed (i.e., with the density
matrix having more than one nonzero eigenvalue, not a pure state
with only one nonzero eigenvalue).  Then one can ask what the typical
information is in the radiation subsystem at various stages of the
black
hole evaporation.

	To control the dimensions of the Hilbert spaces involved,
imagine
forming the black hole from a pure state of radiation in a box.  For
simplicity, suppose the radiation is initially in a superposition of
energy
and angular momentum eigenstates with eigenvalues clustered near
$E$ and 0 respectively (so that the initial state is essentially one
pure
state out of a microcanonical ensemble with zero angular momentum).
Let the box volume $V$ initially give $E^5/V$ large enough in Planck
units,
so that once a black hole forms, it would be semiclassically stable
in the
microcanonical ensemble, with the black hole having more than 4/5 of
the total energy
\cite{Haw76a}.  (One could imagine that the black hole forms
either by having the radiation initially aimed inward to collapse, or
else
by squeezing the box until the radiation suffers Jeans collapse.)

\newpage
	In the spirit of the hypothesis that no information is lost
in black
hole formation and evaporation, assume that the radiation subsystem
has dimension $m \sim e^{s_r}$, where $s_r$ is the thermodynamic
radiation entropy, and the black hole subsystem
has a Hilbert space dimension $n \sim e^{s_h}$, where $s_h = A/4$
is the semiclassical Hawking entropy \cite{Haw74,Haw76a} of a black
hole of area $A$.  Since the
total angular momentum in the box is assumed to be zero, most of the
black hole states would have little rotation and would be nearly
Schwarzschild with mass $M$, so $s_h\simeq 4\pi M^2$.
Assuming the box is much larger than the hole, so $E^3\ll V$, most
of the spacetime within the box will be nearly flat.  If the
radiation is
in semiclassical equilibrium with the black hole of Hawking
temperature
$ \simeq (8\pi M)^{-1}$,
the energy and thermodynamic entropy of the radiation would be
     	\begin{equation}
	E-M \simeq a(8\pi M)^{-4} V,
	\end{equation}
     	\begin{equation}
	s_r \simeq \frac{4}{3} a(8\pi M)^{-3} V,
	\end{equation}
where $a$ is the radiation constant for the species of massless
particles
present, assuming a negligible contribution from massive particles.

	The idea now is that the radiation and the black hole are
subsystems
of a total system of Hilbert space dimension $mn$.  Although the
total
system is in a pure state with density matrix $\rho_{rh} =
\rho_{rh}^2$,
each subsystem is in a mixed state,
     	\begin{equation}
	\rho_r = \mathop{\hbox{tr}}_h \rho_{rh},\;\;
	\rho_h = \mathop{\hbox{tr}}_r \rho_{rh},
	\end{equation}
with a von Neumann or entanglement entropy
     	\begin{equation}
	S_r  = -\mathop{\hbox{tr}}_r (\rho_{r} \ln \rho_{r}) =
	S_h = -\mathop{\hbox{tr}}_h (\rho_{h} \ln \rho_{h})
	\end{equation}
and information (deviation of the entanglement entropy from maximum)
     	\begin{equation}
	I_r = \ln m -S_r \simeq s_r - S_r,\;\; I_h = \ln n - S_h
\simeq s_h - S_h.
	\end{equation}

	We would like to know how much information $I_r$ to expect in
the
radiation at various stages of the black hole evaporation.  Without a
full
quantum analysis, we cannot really answer this question definitely.
However, a reasonable first guess would be that $I_r$ is near the
average
information in a subsystem of dimension $m$ when the total system,
of dimension $mn$, is in a random pure state.  For $m \leq n$
(which is the case for a locally stable black hole in a box), this
average
appears to be \cite{PagEnt}
	\begin{equation}
	I_{m,n}=\ln m + \frac{m-1}{2n}-\sum_{k=n+1}^{mn}\frac{1}{k},
	\end{equation}
and for $1 \ll m \leq n$, it can more definitely be shown to be
\cite{PagEnt}
	\begin{equation}
	I_{m,n} \simeq \frac{m}{2n} \sim e^{s_r - s_h}.
	\end{equation}
For $m \geq n$, the fact that $S_r=S_h$ implies that Eqs. (5)-(7)
give
	\begin{equation}
	I_{m,n}=\ln m - \ln n + I_{n,m}
	= \ln m + \frac{n-1}{2m}-\sum_{k=m+1}^{mn}\frac{1}{k}
	\sim \ln m - \ln n + \frac{n}{2m}.
	\end{equation}
This average information, along with the average entanglement
entropy $S_{m,n} = \ln m - I_{m,n}$, is plotted versus the subsystem
thermodynamic entropy $\ln m$ in Fig. 1 for the case $mn=291600$,
whose 105 integer divisors are taken to be the values of $m$.

\vspace{1.5 cm}
\centerline{\epsfbox{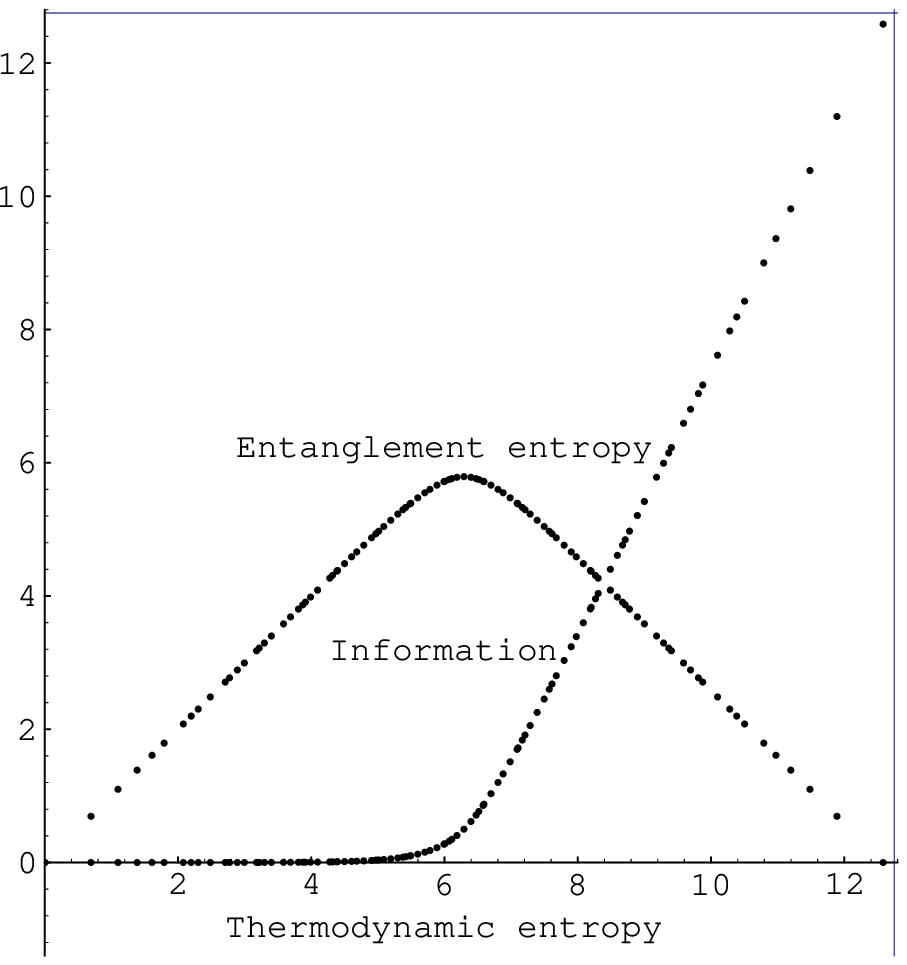}}
\bigskip
\noindent
{\bf FIG.~1.  Average von Neumann or entanglement entropy $S_r  =
-\mathop{\hbox{tr}}_r (\rho_{r} \ln \rho_{r})$ and information
$I_r = \ln m -S_r = s_r - S_r$ of a radiation subsystem of
Hilbert-space dimension $m$ versus its thermodynamic entropy,
here defined to be $s_r = \ln m$.  The radiation is assumed to be
coupled to another subsystem (e.g., a black hole) of dimension $n$
(and hence thermodynamic entropy $s_h = \ln n$), such that the
two subsystems form a combined system in a random pure state in
its product Hilbert space of fixed total dimension $mn$,
here taken to be $2^4 3^6 5^2 = 291600\sim e^{4\pi}$
(about the number of states very na\"{\i}vely expected for
a black hole near the Planck mass).}

\newpage
	Eq. (7) means that for a typical pure quantum state of a
joint
system, the smaller subsystem is very nearly maximally mixed,
showing little sign that the total system is pure.  For example,
when the radiation that has been emitted from a black hole has
a smaller Hilbert-space dimension than that of the hole that remains,
the radiation would typically have very little information in it.
Alternatively, consider the case in which the black hole has emitted
most of its energy, so that the radiation has the larger dimension.
If one then examines only part of the radiation at a time, so that
each part has a smaller dimension than that of the rest of the
system,
one would expect to see in the separate parts only a tiny amount
of the information.  The total information is instead mostly encoded
in the correlations between all the parts.

	As a black hole evaporates, the dimension $n$ of the Hilbert
space of black hole states of energy near the actual (decreasing)
black hole mass decreases, and the effective dimension $m$ of
radiation states macroscopically near the actual radiation state
increases.  Therefore, by Eq. (6)-(8), the expected information in
the radiation (in correlations spread throughout it) also increases.

	For example, a black hole in a box would evaporate
adiabatically as the box is slowly expanded.  This would keep the
total semiclassical entropy, $s \equiv s_r + s_h$, constant.
Under the assumption that the box is much larger than the hole,
which is itself large in Planck units, so $1 \ll E^3 \ll V$, but not
necessarily $V \ll E^5$, one can conveniently parametrize
the adiabatic expansion by the monotonically growing ratio of
the radiation energy to the black hole energy,
	\begin{equation}
	x \equiv \frac{E-M}{M} \simeq \frac{aV}{(8\pi)^4M^5},
	\end{equation}
which must be less than $\simeq 1/4$ for locally stable equilibrium
\cite{Haw76a}.  Then as a function of $x$ and of the energy
$E_0$ when $x$ is negligibly small (which is when $V \ll E^5$),
the parameters of the hole and radiation in
adiabatic equilibrium at constant $s \simeq 4\pi E_0^2$ vary as
	\begin{equation}
	E \simeq E_0(1+x)(1+8x/3)^{-1/2},
	\end{equation}
	\begin{equation}
	M \simeq E(1+x)^{-1} \simeq E_0(1+8x/3)^{-1/2},
	\end{equation}
	\begin{equation}
	T \simeq (8\pi M)^{-1} \simeq (8\pi E_0)^{-1}(1+8x/3)^{1/2},
	\end{equation}
	\begin{equation}
	V \simeq (8\pi)^4 a^{-1} E_0^5 (1+8x/3)^{-5/2} x,
	\end{equation}
	\begin{equation}
	s_r \simeq s\frac{8x}{3+8x} \simeq 4\pi E_0^2
\frac{8x}{3+8x},
	\end{equation}
	\begin{equation}
	s_h \simeq s\frac{3}{3+8x} \simeq 4\pi E_0^2 \frac{3}{3+8x},
	\end{equation}
	\begin{equation}
	I_r \sim \exp(4\pi E_0^2 - 8\pi M^2) \sim
	 \exp(-4\pi E_0^2\frac{3-8x}{3+8x}).
	\end{equation}

\baselineskip 14.5pt
	During the adiabatic evaporation stage,
when the box is expanded slowly and
$x{\;\raise.3ex\hbox{$<$\kern-.75em\lower1ex\hbox{$\sim$}}\;}1/4$,
the rate at which information comes out of the black hole is roughly
	\begin{equation}
	\frac{dI_r}{dt} \sim \exp(-4\pi E_0^2 \frac{3-8x}{3+8x})
	\frac{dx}{dt}.
	\end{equation}
(The prefactor in front of the exponential has been dropped,
since it is less relevant than corrections to the large negative
exponent.)  At the beginning of the black hole emission,
when (in Planck units) $M \simeq E_0 = 1/y$ in terms of the small
perturbative parameter
	\begin{equation}
	y = M_{Planck}/E_0,
	\end{equation}
the initial rate of information outflow appears to be roughly
	\begin{equation}
	\frac{dI}{dt} \sim e^{-4\pi/y^2}.
	\end{equation}
This is not analytic in $y$ at $y=0$, so we would never find it by
an order-by-order (perturbative) analysis in four dimensions
analogous to the one that Giddings and Nelson propose
\cite{GidNel,Gid92} in two dimensions.  Therefore, even if it
can be proved that the information is not emitted at any finite
order of a perturbation series in $y$, it would be consistent
with the most na\"{\i}ve expection of what would happen if the
information were indeed coming out, and so it would not be
evidence against that conservative possibility.

	One may also apply this argument directly to the toy model
\cite{CGHS} of two-\linebreak dimensional dilatonic black holes
actually analyzed by Giddings and Nelson \cite{GidNel,Gid92}.
They propose to rule out a gradual emission of the information
by a perturbative analysis in $1/N$ for large $N$, the number
of minimally coupled scalar fields.

	When the quantum corrections are small so that
the classical equations of \cite{CGHS} provide good
approximations for the various thermodynamical quantities,
rewriting the null coordinates therein as
	\begin{equation}
	x^-=\sqrt{M/\lambda^3}\:u,\;\;x^+=\sqrt{M/\lambda^3}\:v,
	\end{equation}
gives the classical black hole metric \cite{2DBH}
	\begin{equation}
	ds^2=\frac{-\lambda^{-2} du\,dv}{1-uv}
	=-\tanh^2(\lambda r)dt^2+dr^2,
	\end{equation}
where the last expression applies only outside the horizon,
for $u=-\sinh(\lambda r)e^{-\lambda t}<0,\;
v=\sinh(\lambda r)e^{\lambda t}>0$.  This metric depends
only on the cosmological constant $\lambda^2$
(which sets the scale) and is independent of the mass $M$.
Therefore, it is not surprising that the Hawking temperature
at $r=\infty$ (where $g_{tt}=-1$),
	\begin{equation}
	T=\frac{\kappa}{2\pi}
	=\frac{1}{2\pi}\frac{d}{dr}\tanh(\lambda r)|_{r=0}\,
	=\frac{\lambda}{2\pi},
	\end{equation}
is independent of $M$ (in the classical limit).

\baselineskip 15.1pt
	In the classical solution, the mass $M$ occurs only
in the expression for the dilaton,
	\begin{equation}
	e^{-2\phi}=\lambda^{-1}M(1-uv)
	=\lambda^{-1}M\cosh^2(\lambda r).
	\end{equation}
That is,
	\begin{equation}
	M=\lambda e^{-2\phi_H}
	\end{equation}
in terms of the value $\phi_H$ of the dilaton at the horizon,
$uv=0$ or $r=0$.

	The first law then gives the thermodynamic entropy
of the two-dimensional black hole as
	\begin{equation}
	s_h=\int dM/T=2\pi M/\lambda=2\pi e^{-2\phi_H},
	\end{equation}
up to a constant of integration that I shall assume is
negligible when $s_h$ is large.  The quantum-corrected
equations \cite{CGHS} are valid outside the horizon for
	\begin{equation}
	e^{-2\phi_H}
	{\;\raise.3ex\hbox{$>$\kern-.75em\lower1ex\hbox{$\sim$}}\;}
	N/24\;\;
	\mbox{\rm or }\;s_h
	{\;\raise.3ex\hbox{$>$\kern-.75em\lower1ex\hbox{$\sim$}}\;}
	\pi N/12,
	\end{equation}
so the minimum thermodynamic entropy of a
two-dimensional black hole is bounded below by a
constant (of order unity) times the number $N$ of scalar fields,
at least if we stay in the regime where the $1/N$ expansion
is valid.

	If the two-dimensional black hole plus radiation
outside can be treated as a coupled joint quantum
system with unitary evolution, then it would be reasonable
to suppose that the number of quantum states of the
black hole of mass near $M$ would be roughly
	\begin{equation}
	n\sim e^{s_h}
	{\;\raise.3ex\hbox{$>$\kern-.75em\lower1ex\hbox{$\sim$}}\;}
	e^{\pi N/12}.
	\end{equation}
If the black hole has radiated an effective Hilbert-space
dimension $m\sim e^{s_r}<n$ of radiation with
thermodynamic entropy $s_r<s_h$, then the information
in the radiation would be expected to be
	\begin{equation}
	I_r\sim I_{m,n}\sim\frac{m}{2n}\sim e^{s_r-s_h}
	{\;\raise.3ex\hbox{$<$\kern-.75em\lower1ex\hbox{$\sim$}}\;}
	e^{s_r-\pi N/12}=e^{s_r-\pi/(12z)},
	\end{equation}
where now the small parameter is $z=1/N$.

	Thus we see that for a fixed effective dimension
of radiation states (e.g., Hawking evaporation at the
$M$-independent temperature for a fixed time),
the inequalities (26)-(28) imply that the information
expected in the radiation would not be analytic in
$z=1/N$ at $z=0$.  Therefore, even if the information
were coming out in nonthermal corrections throughout
the Hawking radiation, one would not expect to see it
by the order-by-order (perturbative) analysis that
Giddings and Nelson advocate.  That is, even if they
succeed in their goal of proving that the information
does not come out at any finite order of the perturbation,
it would not be a convincing argument that the information
is not actually coming out in a nonperturbative way,
since that seems to be the typical behavior for
a random joint pure state of a black hole plus radiation.

	Giddings has argued \cite{Gidpri} that the $1/N$
expansion should be valid until $s_h$ gets down to be
of order $N$, by which time $s_r$ can be much larger.
That is, now the radiation would be the larger subsystem,
with dimension $m>n$. Then the information in the
radiation would, by Eq. (8), be expected to be
	\begin{equation}
	I_r\sim\ln m - \ln n + \frac{n}{2m}\sim
	s_r - s_h + \frac{1}{2}e^{s_h-s_r},
	\end{equation}
which would not be exponentially small as Eq. (28) gave
for $s_r<s_h$.

	However, to test whether or not the information is there,
one would expect to have to measure most of the $m^2-1$
independent real parameters of the density matrix of the
radiation of Hilbert-space dimension $m$.  This would
presumably require a number of measurements at least of
order $m^2\sim e^{2s_r}>n^2\sim e^{2s_h}
{\;\raise.3ex\hbox{$>$\kern-.75em\lower1ex\hbox{$\sim$}}\;}
e^{\pi N/6}$,
which is thus at least exponentially large in $N$.
Therefore, this task becomes enormously more difficult
at $N$ is increased, despite the greater accuracy of the
prediction of each measurement by the $1/N$ expansion.

	For example, suppose that for large $N$ the rms
error of the prediction for each measurement could in
principle be made smaller than any finite power of $1/N$
in the perturbative expansion.  But when one squares
and sums the errors for the more than $e^{\pi N/6}$
measurements typically necessary to determine the
information, one does not have a result that can be
controlled by making $N$ large.  Therefore, this
perturbative analysis apparently could not say
whether the information is there or not.

	Of course, my argument that the hypothesis of the gradual
emission of the information apparently cannot be disproved by
perturbative analyses does not prove the truth of the hypothesis
either.  We do not even yet know any really plausible
mechanisms for getting the information out from inside what
classically appears to be a black hole.  (One might consider
\cite{Hoo,BekSch,uni2DBH,STU}, but in my opinion it is too early
to say positively that any of them is yet very plausible.)
Another idea, described briefly at the end of my review article
in \cite{Pag93}, is that the information is brought out from near
the center of the black hole to near the (apparent) horizon by
wormholes, threads, tubes, or energy conduits, which can
be described as topologically nontrivial or trivial narrow regions
where the metric and causal structure are much different from
that of the surrounding spacetime, due to quantum fluctuations.
However, there is not space to describe this speculative idea
here in more detail.

	In conclusion, if all the information going into
gravitational
collapse escapes gradually from the apparent black hole, it
would likely come at initially such a slow rate or
be so spread out (requiring so many measurements)
that it could never be found or excluded by a perturbative analysis.
No really plausible nonperturbative mechanisms are known for
bringing out the information, but one can speculate about how it
might conceivably be brought out from behind the apparent horizon.

\newpage
	{\bf Acknowledgments}:  Appreciation is expressed for
the hospitality of the Aspen Center for Physics in Colorado
and of Kip Thorne and Carolee Winstein at their home in Pasadena,
California, where the calculations leading to Eq. (7) were done.
I was grateful to have the opportunity to report on this at
the California Institute of Technology, the
University of California at Santa Barbara, the Journ\'{e}es
Relativistes
'93 at the Universit\'{e} Libre de Bruxelles, the 5th Canadian
Conference on General Relativity and Relativistic Astrophysics
at the University of Waterloo, and the Institute for Theoretical
Physics Conference on Quantum Aspects of Black Holes,
where many people gave useful comments,
including Jacob Bekenstein, Esteban Calzetta,
John Friedman, Gary Gibbons, Steve Giddings, James Hartle,
Jeff Harvey, Stephen Hawking, Justin Hayward, Gerard 't Hooft,
Gary Horowitz, John Preskill, Andy Strominger, Lenny Susskind,
L\'{a}rus Thorlacius, and Kip Thorne.  (See also \cite{Pag93} for
other
people I have communicated with at various times about related
issues.)
Part of this work is summarized in my contributions
for the proceedings of the latter two conferences \cite{Brux,Pag93}.
I am grateful to one referee for pointing out the need
to clarify the relation between the expansion in the
inverse energy $y$ of Eq. (18) and the expansion in
$z=1/N$, and to another referee for suggesting
the inclusion of a figure.
Thanks are also due to William and Barbara Saxton
for their hospitality in State College, Pennsylvania,
where this paper was first written up.
Finally, gratitude is extended to the
Natural Sciences and Engineering Research Council of
Canada, which provided financial research support.

\end{document}